\documentclass{revtex4}

\usepackage{amssymb}
\usepackage{graphicx}

\begin{document}

\title{The effect of pressure on La$_{1.5}$Ca$_{0.5}$CoIrO$_{6}$ re-entrant spin-glass}

\author{L. T. Coutrim}
\affiliation{Instituto de F\'{\i}sica, Universidade Federal de Goi\'{a}s, 74001-970, Goi\^{a}nia, GO, Brazil}

\author{E. M. Bittar}
\affiliation{Centro Brasileiro de Pesquisas F\'{\i}sicas, 22290-180, Rio de Janeiro, RJ, Brazil}

\author{L. Mendon\c{c}a-Ferreira}
\affiliation{CCNH, Universidade Federal do ABC (UFABC), 09210-580, Santo Andr\'{e}, SP, Brazill}

\author{L. Bufai\c{c}al}
\email{lbufaical@ufg.br}
\affiliation{Instituto de F\'{\i}sica, Universidade Federal de Goi\'{a}s, 74001-970, Goi\^{a}nia, GO, Brazil}

\begin{abstract}
La$_{1.5}$Ca$_{0.5}$CoIrO$_{6}$ is a re-entrant spin-glass (SG) that exhibits antiferromagnetic and ferromagnetic couplings at $T\sim95$ and $\sim86$ K, respectively, and at $T\sim25$ K a SG phase emerges. In this work we investigated the effect of hydrostatic pressure ($P$) on La$_{1.5}$Ca$_{0.5}$CoIrO$_{6}$ magnetic properties. By means of magnetization as a function of temperature measurements, carried under different applied pressures and/or magnetic fields, we have found that the freezing temperature of the SG phase exhibits an initial increase followed by a decrease with increasing $P$, and that the maximum $P$ = 7.9 kbar was not sufficient to prevent the formation of the frozen state. Since the ordering temperatures of the antiferromagnetic and ferromagnetic phases are also affected by $P$, we discuss the results here found in terms of changes on the balance between the magnetic phases. 
\end{abstract}

\maketitle

\section{Introduction}

Double-perovskite (DP) compounds, with general formula A$_{2}$BB'O$_{6}$ [A = rare-earth/alkaline-earth, B and B' = transition-metal (TM)] are the subject of great interest since the discovery of high $T_{C}$ ferrimagnetic (FIM) and half-metallic behavior on A$_{2}$FeMoO$_{6}$ \cite{Kim} and A$_{2}$FeReO$_{6}$ \cite{Kobayashi}.Since then, many other interesting physical properties were found in this family of compounds, such as superconductivity \cite{Rubel}, multiferroicity \cite{Azuma} and glassy magnetic behavior \cite{Serrate}, the latter being strongly associated with disorder and competing magnetic interactions \cite{Mydosh}.

Due to the intrinsic anti-site disorder (ASD) usually found at TM-sites on DP compounds, such class of materials frequently presents competing magnetic interactions and frustration, being thus prospective candidates to exhibit spin-glass (SG)-like behavior. On La$_{1.5}$Ca$_{0.5}$CoIrO$_{6}$, the $\sim$9\% of ASD at Co/Ir sites leads to competing antiferromagnetic (AFM) and ferromagnetic (FM) phases emerging at temperatures ($T$) $\sim95$ and $\sim86$ K, respectively. At lower-$T$, a SG-phase appears, whose existence concomitant with the AFM/FM phases makes this a re-entrant SG (RSG) material \cite{PRB}. 

The RSG behavior observed on La$_{1.5}$Ca$_{0.5}$CoIrO$_{6}$ is associated to the frustration of Ir$^{4+}$, caused by the AFM coupling of Co ions. This compound also exhibits compensation temperatures and spontaneous exchange bias effect, which are believed to be strongly related to the delicate balance between the multiple magnetic phases and to the RSG. To better understand the role played by the magnetic glassiness on this compounds physical properties, it is important to study how the magnetic interactions affect the SG phase. For conventional SG materials, a natural step on this investigation would be to change the concentration of the magnetic ions. For instance, for the canonical CuMn and AuFe systems, which refer to the early works on SG, changing the concentration of Mn or Fe affects the distances between the magnetic ions, which in turn alter the exchange magnetic interactions and leads to significant changes on the SG properties \cite{Mydosh,Morgownik,Maartense}. 

In the case of La$_{1.5}$Ca$_{0.5}$CoIrO$_{6}$, doping on the TM sites is not the proper method of investigation because, due to the strong correlation between structural, electronic and magnetic properties observed on such class of materials, it would not be possible to determine which of these parameters would be most affected by the insertion of different TM ions. On the other hand, applying pressure ($P$) is a suitable way to study how the magnetic properties of this SG material would be affected when the distances between the TM ions vary.

In this work we applied hydrostatic pressures on La$_{1.5}$Ca$_{0.5}$CoIrO$_{6}$, up to 7.9 kbar, and investigated its effect on its magnetic properties. We have found that application of $P$ leads to changes on the freezing-$T$ of the SG spins ($T_{f}$). Since the ordering $T$ of the AFM and FM phases are also affected by $P$, we discuss our results in terms of the imbalance between the magnetic phases caused by the approximation of TM ions. For each investigated $P$, we performed $T$-dependent magnetization curves [$M(T)$] curves under different applied magnetic fields ($H$) and observed that, even for the maximum applied $P$, $T_{f}$ varies with $H$ according to the Almeida-Thouless equation \cite{Mydosh,Thouless}, as expected for SG systems. 

\section{Experimental details}

Polycrystalline La$_{1.5}$Ca$_{0.5}$CoIrO$_{6}$ sample was synthesized by solid state reaction in a conventional tubular furnace and air atmosphere, as described elsewhere \cite{JSSC}. Room-$T$ X-ray powder diffraction confirmed the formation of a single phase monoclinic DP, belonging to $P2_1/n$  space group. 

The magnetic properties under pressure ($P<$ 8 kbar) were investigated in a Quantum Design MPMS3 magnetometer. Hydrostatic pressure conditions were achieved on a Cu-Be piston-cylinder type cell. Silicon oil was used as pressure transmitting medium and the $P$ value on the sample region was evaluated from the shift of the superconducting critical temperature of a small lead piece. For each $P$, we measured $M(T)$ curves with at least three different $H$ according to the zero field cooled (ZFC) and field cooled (FC) procedures. We systematically followed the same protocol for the acquisition of any of the curves, \textit{i.e.}, the same $T$- and $H$-sweeping rates and so on.

\section{Results and discussion}

\begin{figure}
\begin{center}
\includegraphics[width=0.5\textwidth]{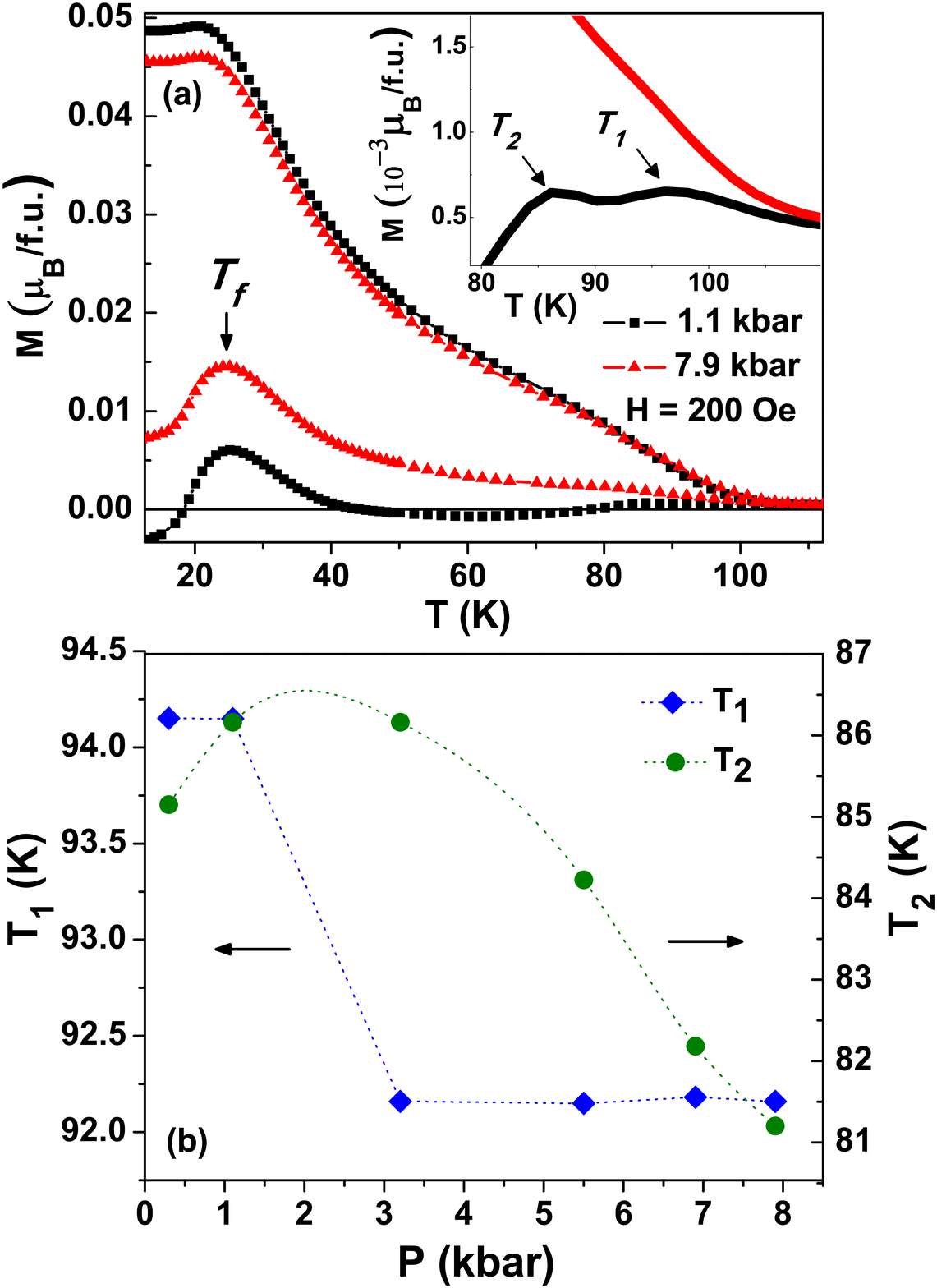}
\end{center}
\caption{(a) ZFC and FC $M(T)$ curves for $P$ = 1.1 k (black squares) and 7.9 kbar (red triangles), at $H$ = 200 Oe. Inset shows a magnified view of the ZFC curves close to 90~K, evidencing two anomalies at $T_{1}$ and $T_{2}$ K. (b) $T_{1}$ and $T_{2}$ $P$-dependence for the $M(T)$ measurements performed at $H$ = 200 Oe. The dashed lines are guides for the eye.}
\label{FigMxT}
\end{figure}

Figure \ref{FigMxT}(a) shows the ZFC and FC curves for $P$ = 1.1 kbar and 7.9 kbar, performed at $H$~=~200~Oe. Both are very similar to that measured at ambient $P$ \cite{PRB}, with the presence of anomalies associated to the emergence of conventional magnetic phases, and a cusp at $T\sim$25 K that is ascribed to the freezing $T$ of the SG phase. The curves obtained for the other pressures (not shown) are also very alike those observed on Fig. \ref{FigMxT}. 

Inset of Fig. \ref{FigMxT}(a)shows a magnified view of the 1.1 kbar ZFC curve close to the region where two anomalies at $T_{1}\simeq94.1$ K and $T_{2}\simeq86.2$ K are evidenced, being these associated to the AFM and FM couplings of the TM ions. This is again in accordance with the results found at ambient $P$ \cite{PRB,JMMM2017}. However, one can note changes for the 7.9 kbar curve in relation to the 1.1 kbar one. Not only the slope of the curve varies, but also the temperatures of the magnetic transitions. This becomes more evident on Fig. \ref{FigMxT}(b), which displays the $T_{1}$ and $T_{2}$ values found for all investigated $P$s. Interestingly, $T_{1}$ and $T_{2}$ vary differently with $P$. While $T_{2}$ firstly increases and subsequently decreases with increasing $P$, $T_{1}$ goes through a great decrease from 1.1 k to 3.2 kbar. These results give clear indicative of the influence of $P$ on the magnetic couplings between the TM ions, which alter the magnitude of the AFM and FM magnetic phases. Since the peculiar magnetic properties observed for this material are related to a delicate balance between the multiple magnetic phases present on it, these changes certainly affects the SG state.

\begin{figure}
\begin{center}
\includegraphics[width=0.5 \textwidth]{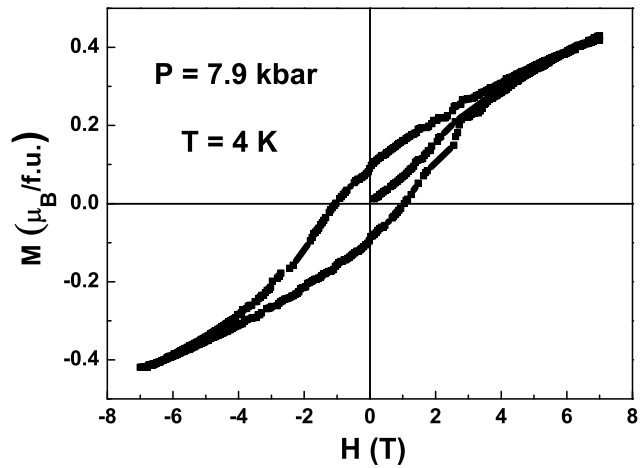}
\end{center}
\caption{$M(H)$ loop obtained under $P$~=~7.9~kbar and at 4~K after ZFC the sample.}
\label{FigMxH}
\end{figure}

For $P$ = 0.3, 1.1 and 7.9 kbar, it was performed magnetization as a function of $H$ [$M(H)$] curves at $T$ = 4 K, with a maximum applied $H$ of 70~kOe, measured after cooling the system in zero $H$. Figure \ref{FigMxH} shows the $M(H)$ loop for $P$ = 7.9 kbar. This curve is very similar to that found for ambient $P$ measurements \cite{PRB}, indicating that even the maximum applied $P$ was not sufficient to suppress the formation of any of the magnetic phases present in the system. 

Here we must stress that all measurements were performed with La$_{1.5}$Ca$_{0.5}$CoIrO$_{6}$ sample together with a Pb sample used as a sensor for $P$. Since it was not possible to measure $M(H)$ curves only for Pb in order to subtract its contribution to the magnetization (\textit{i.e.}, we could not measure only Pb at the same $P$ used to investigate La$_{1.5}$Ca$_{0.5}$CoIrO$_{6}$), and taking into account that the variations on the $M(H)$ curves obtained with different $P$ are visible but small, one could not extract reliable quantitative information concerning the variation of some parameters with $P$, as the remanent magnetization and coercive field. The same argument holds for the $M(T)$ curves, for which it was observed interesting changes on the magnitude of the cusps related to the magnetic transitions. Despite the fact the magnetic signal of our La$_{1.5}$Ca$_{0.5}$CoIrO$_{6}$ sample is of the order of $\sim$10$^{-2}$ emu while the signal of Pb  together with the $P$ cell is of $\sim$10$^{-5}$ emu, one can not state for sure that the variations on the magnitude of magnetic signal are related with changes in La$_{1.5}$Ca$_{0.5}$CoIrO$_{6}$ mainly.

\begin{figure}
\begin{center}
\includegraphics[width=0.5\textwidth]{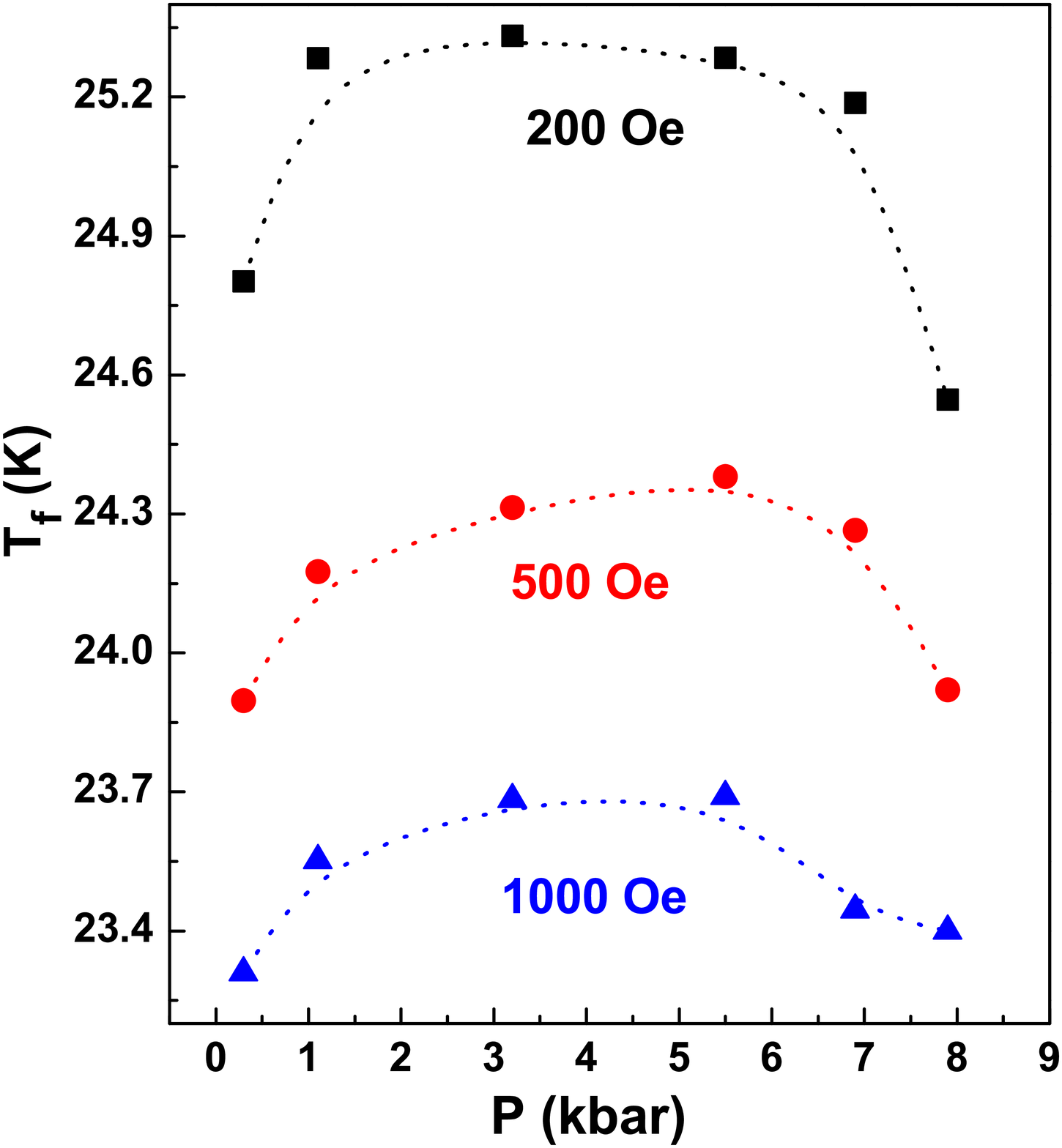}
\end{center}
\caption{Variation of $T_{f}$ with $P$ for three different magnetic fields: 200 Oe (squares), 500 Oe (circles) and 1000 Oe (triangles). The dashed lines are guides for the eye.}
\label{FigTf}
\end{figure}

As described above, application of pressure affects the magnitude of the AFM and FM phases present on La$_{1.5}$Ca$_{0.5}$CoIrO$_{6}$. Since the magnetic properties of the material are known to result from a delicate balance among these phases, the changes driven by $P$ certainly affects the SG phase. Figure \ref{FigTf} shows $T_{f}$ as a function of $P$ for the three applied $H$. As it can be seen, all curves present basically the same trend with an initial increase, followed by a decrease, of $T_{f}$ with increasing $P$. This behaviour resembles those found for the early SG systems such as Au$_{1-x}$Fe$_{x}$ and Eu$_{x}$Sr$_{1-x}$S \cite{Mydosh,Binder}, for which the increase on the concentration of the magnetic ion acts as a chemical pressure, shortening the distance between them. On the ascending branches of the three curves on Fig. \ref{FigTf}, the rate of $T_{f}$ variation with $P$, $dT_{f}$/$d$P, could be estimated to vary from $\sim$50 to $\sim$150 mK/kbar, lying thus within the range expected for SG systems with single spin-spin exchange interactions \cite{Hardebusch}.

The increase and subsequent decrease of $T_{f}$ with $P$ is also similar to the effect of chemical pressure observed on the phase diagram of Fe$_{x}$Mg$_{1-x}$Cl$_{2}$ \cite{Binder,Bertrand}. For this material, the Fe to Mg substitution affects the AFM and FM interactions present on the system, leading to the initial increase of $T_{f}$. On the other hand, the Fe/Mg different ionic radius also leads to structural deformation on doping. One can conjecture an analogous situation for La$_{1.5}$Ca$_{0.5}$CoIrO$_{6}$, where hydrostatic pressure affects the crystal structure as well as the magnetic coupling between the TM ions.

\begin{figure}
\begin{center}
\includegraphics[width=0.5\textwidth]{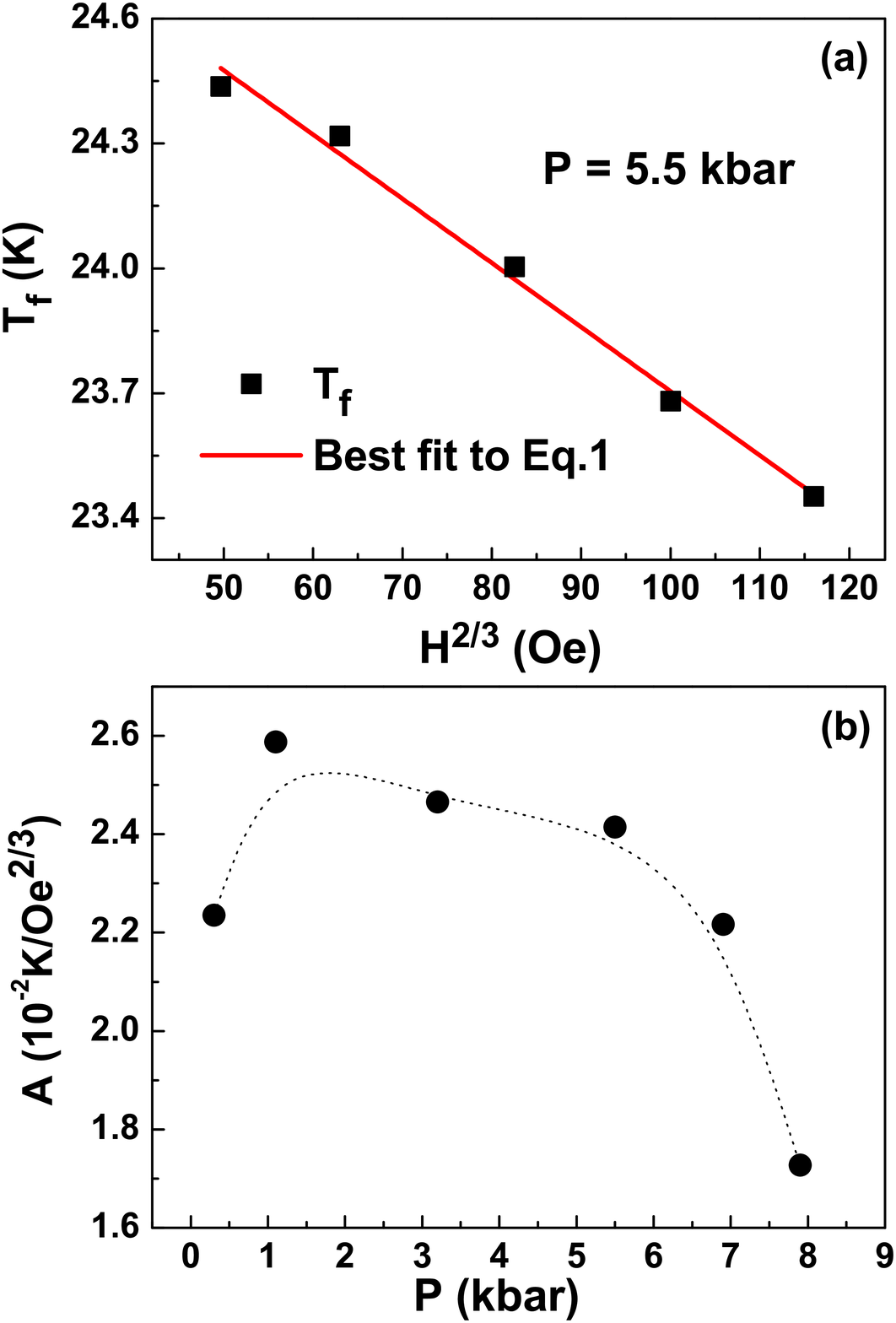}
\end{center}
\caption{(a)Evolution of $T_{f}$ with $H^{2/3}$ obtained from the ZFC $M(T)$ curves for $P$ = 5.5 kbar. The black squares correspond to the experimental results, while the red solid line represent the best fit to Eq. \protect\ref{Eq1}. (b) Evolution of $A$ parameter as a function of $P$, obtained from the AT equation (see text). The dashed line is a guide for the eye.}
\label{FigAT}
\end{figure}

In order to further investigate the effect of $P$ on the magnetic interactions present on La$_{1.5}$Ca$_{0.5}$CoIrO$_{6}$, we have performed for one selected $P$ = 5.5 kbar $M(T)$ measurements with $H$ = 750 and 1250 Oe, besides the 200, 500 and 1000 Oe above described. It was found a monotonic decrease of $T_{f}$ with increasing $H$, as seen on Fig. \ref{FigAT}(a). It can also be noted a reasonable linear dependence of $T_{f}$ with -$H^{2/3}$. This is an expected feature of SG systems, according to Almeida-Thouless theory, which predicts a relation of the form \cite{Mydosh,Thouless,Binder}
\begin{equation}
T_{f} \propto -AH^{2/3},
\label{Eq1}
\end{equation}
where the $A$ parameter is related to the net magnetic interactions between the spins ($J$). Figure \ref{FigAT}(a) displays the best fit of Eq. \ref{Eq1} to the experimental data, where can be observed that the Almeida-Thouless (AT) equation describes the results properly. For the other applied pressures, although the fact they were investigated for only three different $H$, one could also obtain good agreement of the results with Eq. \ref{Eq1}. Thus, one can state that the magnetic glassiness of La$_{1.5}$Ca$_{0.5}$CoIrO$_{6}$ remains even for the largest $P$ here investigated.

Figure \ref{FigAT}(b) shows the evolution of $A$ parameter as a function of $P$. Interestingly, it follows a similar trend to that observed for $T_{f}$, which in turn is related to the evolution of the AFM and FM ordering temperatures as a function of $P$. Bearing in mind that $A$ is directly related to the net coupling between the magnetic ions present in the system, one can state that our results can be understood in terms of the impact of hydrostatic pressure on the exchange interactions among the TM ions, and its consequent influence on the balance between the AFM and FM phases. One can also make an analogy between the curves of Figs. \ref{FigTf} and \ref{FigAT}(b) with the $T$-$J$ phase diagram of Sherrington-Kirkpatrick theory for Ising SG systems \cite{Mydosh,Binder}. On this archetypal phase diagram, the SG phase keeps stable at $T<T_{f}$ (with $T_{f}$ nearly constant) up to a certain critical $J$ value, after which $T_{f}$ continuously decreases. Once again, a relation is made between external pressure and the effect of chemical substitution on canonical SG systems. However, the highest applied pressure in the study here reported was not large enough to prevent the formation of the frozen state on La$_{1.5}$Ca$_{0.5}$CoIrO$_{6}$.

\section{Conclusion}

In conclusion, we have investigated the effect of hydrostatic pressure on the magnetic properties of La$_{1.5}$Ca$_{0.5}$CoIrO$_{6}$. Our results indicate that changes on $T_{f}$ are related to variations of the AFM and FM conventional magnetic couplings, caused by $P$. From $M(T)$ measurements performed under different $H$, we have observed that, even under the effect of $P$, $T_{f}$ evolves with $H$ in accordance to the AT equation, as expected for SG systems. These results show that even the highest $P$ here achieved (7.9~kbar) was not sufficient to prevent the emergence of the SG phase at low-$T$. An analogy can be made between the effect of pressure on La$_{1.5}$Ca$_{0.5}$CoIrO$_{6}$ and the chemical substitution on canonical SG materials, for which the insertion of magnetic ions have the effect of a chemical $P$, shortening the distances between the magnetic ions.

\section*{Acknowledgments}
This work was supported by the Brazilian funding agencies CNPq (Grants No. 470.613/2012-2; 304649/2013-9; 442230/2014-1; 400134/2016-0), FAPERJ (Grant No. 111.382/2013), and FAPESP (Grant No. 2011/19924-2).


\section{References}

\begin{thebibliography}{99}

\bibitem{Kim} T. H. Kim, M. Uehara, S.-W. Cheong, S. Lee, Appl. Phys. Lett. 74 (1999) 1737.

\bibitem{Kobayashi} K. -I. Kobayashi, T. Kimura, Y. Tomioka, H. Sawada, K. Terakura and Y. Tokura, Phys. Rev. B 59, 11159 (1999).

\bibitem{Rubel} M. H. K. Rubel, A. Miura, T. Takei, N. Kumada, M. M. Ali, M. Nagao, S. Watauchi, I. Tanaka, K. Oka, M. Azuma, E. Magome, C. Moriyoshi, Y. Kuroiwa, and A. K. M. A. Islam, Angew. Chem. Int. Ed. 53 (2014) 3599. 

\bibitem{Azuma} Y. Shimakawa, M. Azuma, and N. Ichikawa, Materials 4 (2011) 153.

\bibitem{Serrate} D. Serrate, J. M. De Teresa, and M. R. Ibarra, J. Phys.: Condens. Matter 19 (2007) 023201.

\bibitem{Mydosh} J. A. Mydosh, \textit{Spin Glasses: An Experimental Introduction} (Taylor \& Francis, London, 1993).

\bibitem{PRB} L. T. Coutrim, E. M. Bittar, F. Stavale, F. Garcia, E. Baggio-Saitovitch, M. Abbate, R. J. O. Mossanek, H. P. Martins, D. Tobia, P. G. Pagliuso, and L. Bufai\c{c}al, Phys. Rev. B 93 (2016) 174406.

\bibitem{Morgownik} A. F. J. Morgownik and J. A. Mydosh, Phys. Rev. B 24 (1981) 9.

\bibitem{Maartense} I. Maartense and G. Williams, Phys. Rev. B 17 (1978) 1.

\bibitem{Thouless} J. R. L. Almeida and D. J. Thouless, J. Phys. A 11 (1978) 983.

\bibitem{JSSC} L. T. Coutrim, D. C. Freitas, M. B. Fontes, E. Baggio-Saitovitch, E. M. Bittar, E. Granado, P. G. Pagliuso, and L. Bufaiçal, J. Solid State Chem. 221 (2015) 373.

\bibitem{JMMM2017} L. T. Coutrim, E. M. Bittar, E. Baggio-Saitovitch, and L. Bufaiçal, J. Magn. Magn. Mater. 428 (2017) 70.

\bibitem{Binder} K. Binder and A. P. Young, Rev. Mod. Phys. 58 (1986) 801.

\bibitem{Hardebusch} U. Hardebusch, W. Gerhardt, and J. S. Schilling, Phys. Rev. Lett. 44 (1980) 352.

\bibitem{Bertrand} D. Bertrand, F. Bensamka, A. R. Fert, J. Gelard, J. P. Redoules, and S. Legrand  J. Phys. C: Solid State Phys. 17 (1984) 1725.

\end{thebibliography}
\end{document}